# Secure Platform for Processing Sensitive Data on Shared HPC Systems

Michel Scheerman[1], Narges Zarrabi[1], Martijn Kruiten[1], Maxime Mogé[1], Lykle Voort[1], Annette Langedijk[1], Ruurd Schoonhoven[2], Tom Emery[3]

[1] SURF, Science Park 140, 1098 XG Amsterdam, The Netherlands
[2] Statistics Netherlands, Henri Faasdreef 312, 2492 JP Den Haag, The Netherlands
[3] Erasmus University Rotterdam, Burgemeester Oudlaan 50, 3062 PA Rotterdam, The Netherlands

```
Michel.Scheerman@surf.nl
Narges.Zarrabi@surf.nl
Martijn.Kruiten@surf.nl
Maxime.Moge@surf.nl
Annette.Langedijk@surf.nl
R.Schoonhoven@cbs.nl
Tom@odissei-data.nl
```

**Abstract.** High performance computing clusters operating in shared and batch mode pose challenges for processing sensitive data. In the meantime, the need for secure processing of sensitive data on HPC system is growing. In this work we present a novel method for creating secure computing environments on traditional multi-tenant high-performance computing clusters. Our platform as a service provides a customizable virtualized solution using PCOCC and SLURM to meet strict security requirements without modifying the existing HPC infrastructure. We show how this platform has been used in real-world research applications from different research domains. The solution is scalable by design with low performance overhead and can be generalized for processing sensitive data on shared HPC systems imposing high security criteria.

## 1 Introduction

Shared high-performance computing (HPC) facilities are very popular due to their accessibility and cost-effectiveness [1] [2]. Such HPC systems have proven themselves as an effective solution for running sophisticated codes and parallel processing large volumes of data [3] [4]. Typically, HPC clusters operate in a shared and batch mode, which means that many users can run jobs simultaneously on separate sets of compute nodes in the cluster. The shared nature of these HPC systems however, makes it challenging to impose security requirements for processing highly sensitive or confidential data. With increasing availability of sensitive and confidential data, and data protection laws, such as GDPR in Europe, and HIPAA in the US coming into force, there is a high demand for HPC facilities with a tightly managed security level. In particular in studies where researchers use data made available by third parties, the data owners will impose restrictions on the export of data and results. The data protection requirements can only be met with a fine-grained control over data import and export, in combination with auditing and logging of data access and export. These facilities are typically not present on current HPC systems.

Social science is one research domain where security and privacy of data are extremely important. Scientists in this domain try to understand societal challenges by using sensitive information about human characteristics and behaviors. Specifically, data from surveys or other sources are linked to administrative or genetic data at an individual level, enriching it with diverse and complex data forms that are necessary for scientific research. The advances in high-performance computing enable social scientists to analyze bigger data and run simulations at considerably larger scales than they are traditionally used to. However, social science research has been limited as a consequence of the ethical and legal consideration when handling personal data. The emerging applications of high-performance computing in social sciences requires HPC facilities that take into account the strict security requirements for processing sensitive data. To achieve this goal and to facilitate social scientists in linking, processing and analyzing sensitive data on HPC, three partners from science, academy and government in the Netherlands have come together to make this happen. In the remainder of this section, we briefly introduce the partners involved in this work. The Open Data Infrastructure for Social Science and Economic Innovations (ODISSEI, http://www.odissei-data.nl/en/) is the Dutch national infrastructure for the social sciences [5]. The aim of the infrastructure is to coordinate social science data collection efforts within the Netherlands and ensure their integration with e-infrastructures and alignment with the research aims of the scientific and policy making communities. ODISSEI enables social scientists to bridge out to other fields like humanities, genetics, imaging, epidemiology and computing to better address their research questions.

Statistics Netherlands (Centraal Bureau voor de Statistiek, CBS) holds a wide range of government-based integral administrative microdata on persons, households, firms etcetera in the country [6]. This data is highly confidential but CBS is legally entitled to make it available for research purposes under strict legal terms and security conditions. CBS provides researchers with access to administrative data and the ability to link it to additional external data, within a strictly secured network environment. Remote access to the microdata is available through secure VPN with two-factor authentication. However, the storage capacity and compute power at CBS microdata environment is limited and affects the scale of research studies that can be performed.

SURF maintains and operates the Dutch national HPC infrastructure for academic and industrial use. SURF's supercomputer Cartesius consists of 2000 multi-core compute nodes, with a total theoretical peak performance of 1.843 Pflop/s, connected with Infiniband for high-speed, low-latency communication and storage access. A broad selection of pre-installed and maintained software, tuned for performance is available to facilitate users from all fields. Cartesius offers multiple Terabytes of storage for projects, making data-intensive research possible. For academic users, the security level offered on Cartesius and typical public supercomputers suffices. However, the service is not meant for storing and processing sensitive data that require stricter security measures.

To facilitate cutting edge research, a platform was needed to enable access, linkage and processing of such sensitive data in large scale on HPC, while meeting the security requirements as set by e.g. CBS-act. In what follows we describe a platform that we designed and implemented that can offer tailor-made security to meet the user's requirements, while leveraging the power of a supercomputer.

## 2   Platform as a service

We have developed a "Platform as a Service" called the ODISSEI Secure Supercomputer (OSSC), which brings together three crucial components. Firstly, the platform provides researchers with secure remote access to highly sensitive data such as depersonalized administrative data of the Dutch population, held by CBS. Secondly, scientists can import additional external data from surveys or other data sources and link these to administrative data of CBS while maintaining the privacy of the data. Finally, the platform is situated within a high-performance computing environment at SURF and facilitates large scale processing of sensitive data. We built upon the work presented in [7] and [8], which was built with one use case in mind. We evaluated the system with the end users, and took into consideration the user feedback for simpler access and better parallel performance.

The OSSC platform is based on a virtualized private cluster, where users can use the computing resources and software offered on the Cartesius supercomputer; but with a level of security that meets the requirements set by the data owner. The virtual clusters are automatically provisioned using PCOCC (Private Cloud on a Compute Cluster), developed by CEA [9]. PCOCC is a middleware that facilitates building private clouds on existing publicly accessible clusters. PCOCC provisions virtual clusters using Linux host virtualization technology using KVM/QEMU, and uses Open vSwitch for network virtualization. PCOCC also offers integration with the widely used open source batch scheduler SLURM which is used for deployment. On Cartesius, the PCOCC virtual clusters are launched as SLURM jobs. PCOCC then provisions the virtual clusters on the allocated hardware, and stores technical information about the clusters in an etcd central key-value store. Subsequently, virtual machines are installed with cloud images, and initiated with cloud-init files. Following the initiation, the configuration management is done with SaltStack.

In OSSC, each research project runs in a separate virtual environment with its own virtual network layer and Infiniband partition, and no network traffic can be routed to any other environment, nor is there access possible to any internal or external host other than those that are strictly needed for proper functioning of the system (e.g. the fileserver). The storage level security policy implies that the data is stored on a dedicated Lustre filesystem with standard unix permissions and filesystem logging. Communication with the filesystem is done over InfiniBand in an additional InfiniBand partition where only the fileserver has full membership, preventing inter node communication within this partition. Moreover, users can only access the environment through a VPN and a Citrix system run by CBS. The VPN settings enforce that no other routes can exist on the client system other than the VPN routes. Copy and paste are disabled by the Citrix environment. Data cannot leave the environment before it has been checked by CBS staff for disclosure risks.

## 3      OSSC Architecture

The OSSC platform is composed of a work environment, a compute environment as a secure sandbox, a management environment and a data manager environment, all running as virtual clusters on the existing HPC system. All environments consist of the same building blocks and only differ in their function and configuration. Environments are separated either to have a sandboxed environment, or to make it possible to launch and cancel them independently, or both. Fig. 1 illustrates the OSSC architecture and its main components which are described in this section.

### 3.1      Work environment

A work environment consists of one virtual machine with only limited resources in terms of cores and memory. Every project has its own work environment. The main reason to implement a work environment was to give users permanent access to the platform during their project, to prepare or analyse their data. An HPC system, such as Cartesius, typically includes one or more interactive hosts, where users can login and prepare their data and their batch jobs. The batch jobs are then scheduled on demand, using compute resources only when needed. The OSSC platform, being a batch job in itself, claims resources for both the login and compute nodes for as long as the environment is running. This is less efficient when constant access to the data is required, with little or no compute requirements. Thus, a solution was required where a compute and a work environment could be launched separately, with compute only being available when needed. In order to be able to launch jobs from the work environment to be run on the compute environment, the environments need to be connected on the level of the overlay network. Currently PCOCC does not support the combining of two separate PCOCC SLURM jobs into one virtual environment, therefore, an in-house patch and a script have been created that provide this functionality.

   Multiple of such work environments can run on a single physical node. This environment has a wide selection of available software and a SLURM instance to submit jobs to. All relevant filesystems are mounted, and all relevant software is made available. The work environment is also the endpoint for the VPN tunnel from the CBS environment to SURF. Users can edit files, submit SLURM jobs and use available software in the work environment. The SLURM jobs can either be submitted to a (future) compute environment, or can be run directly on the work environment for testing purposes or small tasks that do not require much compute power. The work environment runs for the duration of the project and will only be destroyed after the project is finished.

### 3.2 Compute environment as a secure sandbox

A compute environment consists of one or more virtual machines which are added to an existing work environment. Unlike the work environment each virtual machine occupies a full node to extract maximum performance out of the hardware. The compute environment is launched upon request for as long as compute is required. The compute environment runs all the SLURM jobs that are submitted by the user in the work environment. After the data processing tasks are finished the sandbox is destroyed and all data that is not stored on the Lustre filesystem is cleared.

### 3.3 Management environment

**Configuration Management.** The management environment is used to configure all other environments and is also a PCOCC virtual cluster. The configuration management is done using SaltStack, with the Salt master running in the management environment. The Salt configuration is kept in a private, self-hosted git repository, and is applied to the other environments during initialization. The Salt master configures the operating system, creates users and installs software. Finally, rules are added to the iptables firewall, and the virtual cluster is no longer able to connect to the Salt Master, nor to any external location.

**Secure logging.** Within the work and compute environments, secure logging is required of login attempts, data access and Linux syslogs. The authentication and system logs are sent to a separate syslog server, which is a separate virtual machine running inside of the management environment. Data access is logged in detail by the Lustre filesystem and is not dependent on any OSSC environment

### 3.4 Data manager environment

An optional data management environment is used in case two data sources need to be combined by a trusted third party (TTP). The data management environment is comparable in setup to the work environment, but is used by the data manager (e.g. CBS) to upload data to a secure location that cannot be accessed by the project members. Likewise, the project members can move their data to a specified location from within the work environment, which cannot be accessed by the data manager. The TTP (e.g. SURF) then processes this data before finally moving it to the project directory. There is typically only one data manager environment per data owner (e.g. CBS), and it's only launched when the data manager needs to upload or access data.

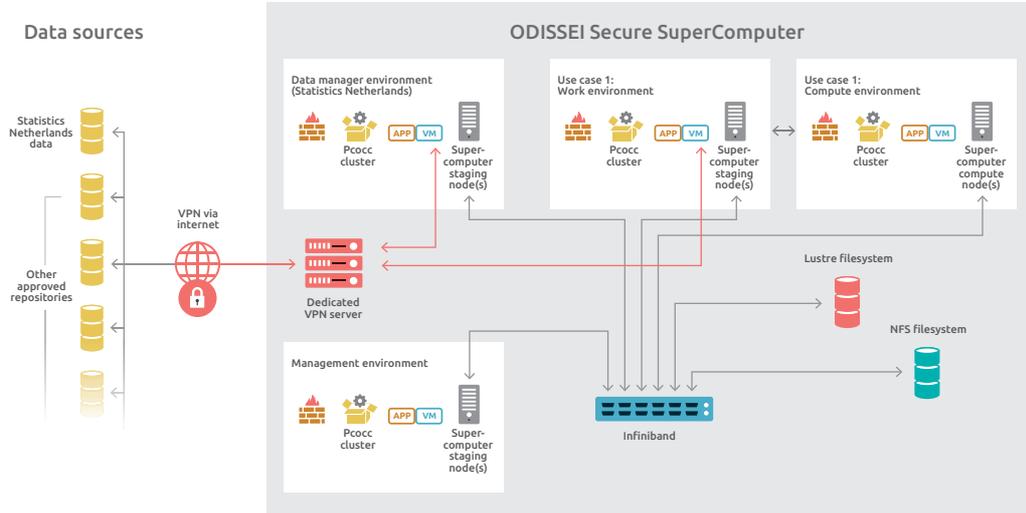

**Fig. 1.** A OSSC Architecture, including work environment virtual cluster for researchers to preprocess data, virtual compute environment for processing sensitive data, virtual management environment for configuration management and secure logging, and a virtual data manager environment for the trusted-third party process.

## 4   Performance Benchmarks

From the secure sandbox, users can allocate any computing resource normally available on the supercomputer via the SLURM batch scheduler and submit jobs using a second instance of the batch scheduler deployed in the secure sandbox. The only performance drawback in this setup is the overhead of the virtualization layer with KVM/QEMU and is to our opinion low enough to allow compute capability similar to that of the normal supercomputer. SR-IOV is used for direct access to the InfiniBand network cards of the host system. Three type of benchmarks were used to measure the overhead of the virtualization layer in OSSC compared to the normal environment on the Cartesius supercomputer: 1) The Triad test from STREAM 10 measuring sustained memory bandwidth on a single node. 2) HPCG 11, which is more computationally intensive and typically used as a metric to measure performances of HPC systems, 3) the Latency and AllReduce tests from the OSU benchmark [12], that measures the latency for point to point and collective MPI communication, and 4) the IOR benchmark [13] to evaluate the performance of the parallel file system.

We ran the benchmarks on Intel Broadwell nodes with 32 E5-2697A v4 CPUs, using GCC compiler v4.8.5 and OpenMPI 4.0.3 with the UCX communication framework. Since the chosen benchmarks use different metrics (MB/s, GFlop/s, s), an efficiency metric was derived by computing the ratio between the values on the OSSC and on the normal environment. A value of 1 means no overhead, a value of 0.95 means an overhead of 5% on the OSSC, and a value greater than 1 means better performances on the OSSC than on the normal Cartesius environment. The benchmark results were derived from the average performance of 5 runs of each benchmark. Table 1 shows that the OSSC has a small performance penalty due to the virtualization layer with efficiencies of 0.92 and above for almost all benchmarks. This performance penalty is considered low enough to consider the performance and scalability similar to that of the normal supercomputer environment. Only the IOR Write benchmark shows a larger penalty, with an efficiency from 0.98 on a single node down to 0.55 on 8 nodes.

**Table 1.** Efficiency of benchmarks on OSSC against normal environment on 32-cores Intel Broadwell nodes of Cartesius Supercomputer.

| Benchmark | | Efficiency |
|---|---|---|
| STREAM Triad – array size = 2G elements – 32 OpenMP threads | | 0.98 |
| HPCG | 1 node / 32 MPI processes | 0.96 |
| | 2 nodes / 64 MPI processes | 0.97 |
| | 4 nodes / 128 MPI processes | 0.97 |
| | 8 nodes / 256 MPI processes | 0.96 |
| OSU Latency – | 1 node – 0 byte message between 2 sockets | 0.95 |
| | 2 nodes – 0 byte message between 2 nodes | 0.92 |
| OSU AllReduce | 1 node / 32 MPI processes | 0.95 |
| | 2 nodes / 64 MPI processes | 0.98 |
| | 4 nodes / 128 MPI processes | 0.96 |
| | 8 nodes / 256 MPI processes | 0.92 |
| IOR Write– using POSIX API for I/O | 1 node / 16 MPI processes | 0.98 |
| | 2 nodes / 32 MPI processes | 0.65 |
| | 4 nodes / 64 MPI processes | 0.63 |
| | 8 nodes / 128 MPI processes | 0.55 |
| IOR Read – using POSIX API for I/O | 1 node / 16 MPI processes | 1.01 |
| | 2 nodes / 32 MPI processes | 0.96 |
| | 4 nodes / 64 MPI processes | 0.98 |
| | 8 nodes / 128 MPI processes | 0.94 |

The OSSC has performance comparable to the Cartesius supercomputer in terms of arithmetic operations (Flops), intra and inter node communications, scalability and memory bandwidth. However, the write performance to the parallel file system (Lustre) measured with the IOR Write benchmark are hampered by the virtual environment. Further investigation is required to understand and fix this. However the maximum write speed measured with IOR Write on 8 nodes on the OSSC is 11622 MB/s, which allows the vast majority of applications to run efficiently. Although we did not test with larges virtual clusters, we have no reason to believe that the scaling of user's applications will be any worse than what can be extrapolated from the performance of the HPCG benchmark, which is the most representative of real use cases of a supercomputer, in Table 1 since the virtual machines have direct access to the Infiniband cards. This makes the OSSC capable of leveraging the power of a supercomputer.

## 5      Secure Access and data transfer

Users can access the sensitive data through the regular 'remote access' microdata environment at CBS. The user's identity is verified through a multi-factor identification system using both a personal token and a code sent via text message. From the remote access environment, data can be copied to OSSC through a dedicated VPN connection between CBS and SURF, seamlessly and securely integrating the HPC cluster into their own private network. Stringent automated security controls make sure this VPN is the only path for sensitive data to leave the cluster. Thus, we have a centralized secure data enclave where the data are being uploaded for analysis. After the analysis is done, the output of the research is checked for non-disclosure, before being sent to the user by email. When the project is finished, all the data will be wiped out from the platform.

## 6      Research Applications

The research applications of the OSSC are broad, but all share the common challenge of analysis with high computational demands, combined with strict control over data access and availability. The first type of applications are projects where the data is currently available in secure environments, like at CBS Microdata Services, but the computational resources available are insufficient. The second type are projects where the OSSC, besides scaling up the compute power, is a common middle ground for linking sensitive data from different parties.

The first instance of research project which makes use of the OSSC, is one that requires increased computational capacity. Social data has been historically small and simple, being largely based on surveys of necessarily limited population size. Computational Social Science has been increasing in recent years because of the increasing availability of administrative data, data collected via the web and due to the increasing number of secure computing facilities which allow for analysis of linking such data at individual record level. In the Netherlands, CBS has been providing this as a service for nearly 20 years. This allows for researchers, subject to security and data protection measures, to access the data via a secure environment. The data in these environments is exceptionally rich and linkable. The computational challenges encountered by researchers in this environment are often associated with the complexity of the data rather than their mere size. Therefore, to understand the complex interdependencies and associations within the data, significant computational resources are required that are unavailable within the existing environment. For example, in analyzing geospatial distributions it is computationally straightforward to calculate small-area estimates, even if there are hundreds of thousands of areas. Computational demands however become extensive if the associations between neighboring areas are examined. Using the OSSC facility, researchers at Delft University were able to calculate small area estimates of entropy in the distribution of non-western migrants [14]. This took four months to calculate in the standard secure environment at Statistics Netherlands but just a week on 24 physical nodes within the OSSC.

Social data's complexity and independence is what makes it social. Societies consist of complex interdependencies between units. Computational capacity is therefore key to a better understanding of social behavior. For example, social scientists have long been aware of the importance of relationships between people, but this is notoriously difficult to collect data on. Using the OSSC, a team of researchers at Statistics Netherlands have created an integral network of Dutch residents in which individuals are linked by work, residential, family or educational connections. This has created a datafile with all 17 million residents of the Netherlands and the 0.8 billion links that exist between them [15]. Fig. 2 shows a synthetic network of a single individual based on observed demographic networks which was obtained through this research study. Using this framework can allow social scientists to better understand how behavior and outcomes spread throughout society, the primary vectors of transmission and the underlying social network structure of the country. Such a framework is a fundamental reframing of sociological operationalizations and a new data paradigm for the entire field.

The second type of project dealt with in the OSSC are those that focus on the linkage of multiple sensitive datasets. In biomedical research the inclination is for the data to remain securely stored in silos and for analytical procedures to be sent to each silo and extract standardized results that can then be aggregated. This is sufficient where the primary analytical requirement is looking at associations of variables within a single data subject (e.g. whether an individual smokes and then whether that same individual subsequently developed cancer). These are then aggregated for the population. Bringing the various data sources together would only serve to append research units to a single data frame. For example, one can study the association between smoking and cancer of individuals in many hospitals without bringing the data together. The OSSC by contrast allows for linkage at the level of individuals, collating multiple records for a single research unit (individuals) and linking them together. For example, where hospitals have detailed data on individual's health records and the statistical office has an individuals' detailed employment record. To examine the association between health and employment history requires the data to be brought together. This is difficult to securely implement using a distributed computational design. The capacity of the OSSC to bring multiple data sources together enables researchers to significantly enhance the range of social and economic data that can be linked with bio-medical and health related data. For example, in an analysis conducted using the OSSC, a team of researchers at the Netherlands Twin Register conducted a Genome-Wide Association Study (GWAS) to identify potential genetic associations with usage of the health care use [16]. A particular (also computational) challenge is to disentangle genetic from environmental factors using dedicated statistical models.

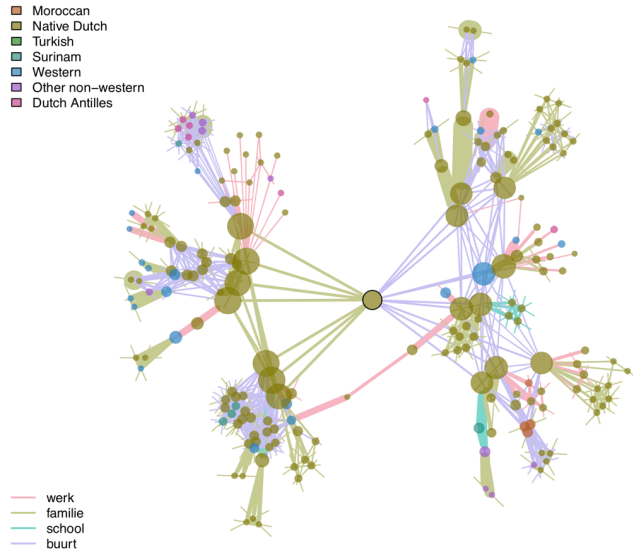

**Fig. 2.** A synthetic example network of a single individual by ethnicity based on the work, family, school and neighborhood connections.

## 7  Discussion

Shared HPC systems are being used by different academic, research and industrial organizations in many fields to process a diverse range of data. The area of privacy sensitive and confidential data is also growing and there is clear need for HPC systems that comply with the policies and security requirements for hosting and processing such data. There are services that offer high performance computing for sensitive data such as, the *TSD Services for Sensitive Data*, developed by the University of Oslo [18]. Their services are based on dedicated infrastructure with a limited scale. Another initiative is the *HPCrypt Data Protection System*, developed by the Lawrence Livermore National Laboratory [19]. This system runs on existing supercomputers and can meet all security requirements that this project aims to meet. The software is, however, proprietary. In this work, we developed a secure platform that makes it possible to process sensitive data on a shared HPC system as a cost effective and scalable solution.

The development of the platform started by a proof of concept with only one research use case to show that the solution and the design concept is feasible. Then we moved to a pilot phase where we scaled the solution for more than one use case. We also did an extensive requirements analysis on the legal terms and security aspects, to understand what is needed for hosting and processing sensitive data on our infrastructure. Those requirements were then translated into technical requirements to be implemented on the platform. Finally, we conducted a penetration test by a security audit company to make sure that the platform satisfies the required level of security. After successfully passing the penetration test, three research use cases ran in the platform and achieved outstanding results, explained in the section Research Applications.

To achieve the collaboration between Statistics Netherlands (CBS) and SURF and make it possible to transfer, store and process sensitive administrative data of CBS on SURF's HPC infrastructure for research purposes, several legal barriers had to be overcome, given not only GDPR(general data protection regulation in EU law) but also CBS Act which among others defines protection of privacy of persons and companies as the very basis for Statistics Netherlands's mission and mere existence. The principal requirement in CBS Act is that no data may ever leave CBS from which information on individual persons, companies etcetera may possibly be derived; this has to be guaranteed under all conditions. Thus, users are not allowed the opportunity to import or export anything into or from their OSSC workspace, nor may external systems be able to get access to that workspace, during any part of the process. Legally, contracts controlling the formal responsibilities and liabilities of CBS and SURF respectively were combined with confidentiality statements signed by SURF data managers and their supervisors. In terms of data protection, pseudo-anonymization methods used as a standard in CBS microdata facility are equally applied to data transferred to OSSC. Access to the secured environment is through a two-factor authentication procedure. Transfer of sensitive data is done through a dedicated site-to-site VPN connection between CBS and SURF. Finally, the analysis results are made available to the researchers outside the secure environment only after checking for disclosure risks by CBS staff.

We are moving towards a production-ready version of the platform by having automated most of the manual steps for setting up the virtual environment and creating a scalable and robust solution. What makes OSSC different from standard SLURM systems is that OSSC environments run in virtual machines run by a non-privileged user with their own virtual network layer. There is no connection possible to external hosts on the internet, nor to internal hosts that are not strictly necessary. Any software that is being used is from our own software stack. Any data that is being used is stored on a dedicated Lustre filesystem and cannot be copied to or from the client machine. Communication with the filesystem is done over Infiniband in a separate Infiniband partition where only the fileserver has full membership, preventing inter node communication within this partition. These measures make the attack surface much smaller than on a general HPC system. In the future we would like to implement more stringent security features such as storage encryption. This can either be a physical shared filesystem as we have now, or a virtual layer on top of another filesystem. For example, we could use encrypted images, which are mounted and shared inside of the virtual cluster by a virtual fileserver, with logging happening inside of the environment. Having access to on-demand compute was one of the main user requests during the pilot. The current procedure in OSSC to reserve compute nodes is still manual and requires the user to contact us by email. In the future, we want to make the access more efficient by developing a portal that users can ask for on-demand compute resources in the secure platform. The availability of the environment was another main improvement point based on the user experience. During the pilot phase, the availability of the platform, running as a job itself, was limited to maximum 5 days. The reason for this was an internal policy of the Cartesius supercomputer, which limits the lifespan of job to maximum 5 days. For the production platform, we developed the work environment which consists of a lightweight virtual machine on shared resources. The compute environment is only added on demand. The main reason for implementing the work environment was that researchers needed permanent (24/7) access to the platform during their project. By setup of a lightweight virtual machines for this purpose we meet the requirements of our users at a low cost.

Statistics Netherlands (CBS) is one of the frontrunners among National Statistical Institutes (NSI's) offering large scale secure access to their microdata for research institutes, together with e.g. the Scandinavian countries and Anglo-Saxon countries like Canada, Australia and New-Zealand. To the best of our knowledge none of these NSI's have realized a collaboration with a national supercomputing center as described here. Apart from legal contracts regulating SURF receiving CBS's confidential data in their domains and covering access of SURF's data managers to the data, the project could only become successful given that the data could be processed in digital environments completely isolated from the outside world. This was achieved on a shared HPC system through the OSSC platform.

From the point of research, social and particularly biomedical research widely uses HPC systems including studies with sensitive variables. A prominent example is the field of genome wide associations studies (GWAS) correlating DNA profiles with behavioral or health variables collected from the same subjects. Such methods have become the basis of a substantial part of our understanding of the role of specific genes for human life. However, OSSC is the first system to our knowledge that allows linkage of e.g. DNA profiles to register data from government-based administrations, thus offering a major enrichment to the behavioral data that can be linked and of the scale those data are available. A specific feature in the GWAS project was that SURF took up the role of trusted third party: by an intricate series of steps in pseudo-anonymizing the data it was guaranteed that linkages could be made while ensuring that the persons' identity in the CBS microdata was not revealed to the researchers while at the same time CBS did not have access to the DNA data themselves. In the other project that was briefly described we showed that social network studies of the integral population of a small or medium size country come into reach with this facility, enabling advanced studies of social segregation along lines of e.g. income, education level of migration background; all prominent challenges in many modern societies.

The Dutch government stimulates Open Science and currently the Dutch scientific community is working hard to adopt the principles [16]. For all educational institutions and research domains, open access publishing and optimal reuse of research data is the norm in 2020. The OSSC platform promotes open science by allowing for the increased linkage and interoperability of sensitive data. Such initiatives are becoming more and more important for research communities. This has far reaching implications for social science research for example and its usage of HPC facilities in the analysis of complex linked data. We are also planning to integrate a secured storage for long-term preservation of the sensitive data in the OSSC platform. Such an archiving system facilitates long-term preservation and therefore, reuse of the data within OSSC, especially data that are linked or enriched through a trusted third party process. The rules for adding metadata to the datasets, as well as regulations how to obtain access are under development in cooperation with the researchers from the ODISSEI community.

## 8      Conclusions and Future work

In this work we have addressed the problem of processing sensitive data on shared HPC systems. This work, to the best of our knowledge, is the first attempt to be able to store and process sensitive data on shared HPC systems without modifying the existing infrastructure. As future work, we are looking into storage encryption and also integrating dedicated filesystems for short-term and long-term storages, which make it more flexible to implement security features. We also want to improve the user experience by automating the procedure for reserving compute nodes and by integrating visual and graphical applications for analysing data such as visual R studio or Jupyter notebooks. Possibility to deploy the platform on other HPC system is another future work to be explored. Currently, we use open source software and tools such as PCOCC and SLURM as building blocks for developing the platform. But to build the platform as a service a lot of configurations are set on top of these tools which are specific to our HPC system, Cartesius, and not with other systems in mind. In future we are aiming to modify and generalize these configurations and patches to be able to deploy the platform on any other HPC system and eventually make the code publicly available. By the time of writing this article we are in contact with a security audit company to plan another penetration test on the production platform before facilitating new research use cases.